\documentclass[a4paper]{jpconf}
\usepackage{graphicx}
\begin{document}
\title{Open charm scenarios}

\author{Qiang Zhao$^{1,2}$}

\address{$^1$ Institute of High Energy Physics, Chinese Academy of
Sciences, Beijing 100049, P.R. China}

\address{$^2$ Theoretical Physics Center for Science Facilities, Chinese Academy of Sciences,
Beijing 100049, P.R. China}

\ead{zhaoq@ihep.ac.cn}

\begin{abstract}
We discuss possibilities of identifying open charm effects in direct
production processes, and propose that direct evidence for the open
charm effects can be found in $e^+ e^-\to J/\psi\pi^0$. A unique
feature with this process is that the $D\bar{D^*}+c.c.$ open channel
is located in a relatively isolated energy, i.e. $\sim 3.876$ GeV,
which is sufficiently far away from the known charmonia $\psi(3770)$
and $\psi(4040)$. Due to the dominance of the isospin-0 component at
the charmonium energy region, an enhanced model-independent cusp
effect between the thresholds of $D^0\bar{D^{*0}}+c.c.$ and $D^+
D^{*-}+c.c.$ can be highlighted. An energy scan over this energy
region in the $e^+e^-$ annihilation reaction can help us to
understand the nature of $X(3900)$ recently observed by Belle
Collaboration in $e^+ e^-\to D\bar{D}+c.c.$, and establish the open
charm effects as an important non-perturbative mechanism in the
charmonium energy region.

\end{abstract}

\section{Background}

Our interests in the study of hadron spectroscopy are strongly
motivated by the speculation of exotic hadrons of which the very
existence is implicated by QCD. During the past decades, the search
for exotic hadrons has been one of the most important physics goals
at high-energy experimental facilities, such as Tevatron (CDF and
D$\emptyset$), B-factories (Belle and BaBar) and electron storage
rings at CLEO and BES-III. Although confirmed evidence for exotics
is still lacking, experimental measurements have brought a lot of
surprises during the past few years. In the charmonium sector, a
number of new resonance-like signals have been observed by the
B-factories~\cite{Olsen:2009ys}, and some of them exhibit peculiar
decay modes which could be an indication for their being exotic
hadrons~\cite{Voloshin:2007dx,Eichten:2007qx,Brambilla:2010cs}.
Meanwhile, it should also be cautioned that some of those
enhancements observed in experiment may not be genuine resonances.
To identify those non-resonant structures and understand the
underlying dynamics for their production and decays are of great
importance for the study of hadron spectroscopy and gaining insights
into the strong QCD.

The most updated summary of those newly-discovered resonance-like
states can be found in the recent review
article~\cite{Brambilla:2010cs}, among which the vector
charmonium-like states are of our interests here. Firstly, the
vector charmonium spectrum is well established in both theory and
experiment in the low mass region. It has been a success of the
potential quark model that the light charmonium states can be well
described by the one-gluon exchange and linear confinement
potentials~\cite{Appelquist:1974yr,Godfrey:1985xj,Barnes:2005pb}. So
far, nearly all the light charmonia below the $D\bar{D}$ open
threshold have been discovered in
experiment~\cite{Nakamura:2010zzi}. The measured masses are in good
agreement with the quark model predictions which suggests that the
massive charm quark is a good effective degrees of freedom in the
description of the charmonium spectrum. Secondly, an experimental
advantage with the vector charmonium states is that they can be
directly produced in $e^+e^-$ annihilations. This would allow a
direct measurement of these states in experiment. As a consequence,
the data would impose an immediate test of theoretical calculations.

Restricted to the vector charmonium sector, the experimental
observations have turned out to be surprising. The first unexpected
result was the observation of the charmonium-like vector $Y(4260)$
in its decay into $J/\psi\pi\pi$ by BaBar in the initial state
radiation (ISR) process. It was later confirmed by Belle and CLEO.
Meanwhile, several other charmonium-like vector states, $Y(4008)$,
$Y(4360)$, and $Y(4660)$, were observed in the same ISR process and
the same $J/\psi\pi\pi$ invariant spectrum. There have been
tremendous theoretical interpretations for these states in the
literature. Some of those states are proposed to be exotic
candidates for which a detailed review can be found in
Ref.~\cite{Brambilla:2010cs}.

Our focus in this proceeding is not on those above mentioned
resonant structures. Instead, we are interested in an enhancement,
the so-called $X(3900)$~\footnote{This structure is named as
$G(3900)$ in Ref.~\cite{Brambilla:2010cs}.}, observed in the ISR
process by Belle in $e^+e^-\to
\gamma(D\bar{D})$~\cite{Pakhlova:2008zza}. The peak position of this
enhancement is located at $3943\pm 21$ MeV with a width of $52\pm
11$ MeV. Several peculiar aspects make this structure extremely
interesting and deserves to be studied in detail: i) In the vicinity
of the $X(3900)$ mass region, there is no vector charmonium state
predicted in the potential quark model. One notices that the nearby
quark model states are $\psi(3770)(1D)$ and $\psi(4040)(3S)$ which
fit well in the $c\bar{c}$ spectrum. This feature immediately
implies that this structure cannot be easily accommodated by the
quark model classifications. ii) A natural speculation is that such
an enhancement is due to the $D\bar{D^*}+c.c.$ open threshold as
anticipated by Eichten {\it et al.}~\cite{Eichten:1979ms} long time
ago. However, theoretical description of the open threshold effects
has suffered a lot from model-dependence. Thus, solid evidence that
confirms the dynamical origin of $X(3900)$ is still desired.

In this proceeding, we report our recent progress on the study of
the open threshold effects in $e^+ e^-\to J/\psi\pi^0$, $J/\psi\eta$
and $\phi\eta_c$. We will show that model-independent evidence can
be reached for the $D\bar{D^*}+c.c.$ open threshold in $e^+ e^-\to
J/\psi\pi^0$. As a consequence, it would be possible to stab the
$X(3900)$ as the first evidence for the open threshold enhancement
observed in experiment.

\section{Identify the open threshold effects}

\subsection{Evidence for the $D\bar{D^*}+c.c.$ open threshold effects in $e^+e^-\to D\bar{D}$}

It should be mentioned that the Belle Collaboration also reported
their measurement of the cross section lineshape of $e^+ e^-\to
D\bar{D^*}+c.c.$~\cite{Abe:2006fj}. The same channels were also
measured by the BaBar Collaboration~\cite{:2009xs} which appeared to
be consistent with the Belle data. In
Refs.~\cite{Abe:2006fj,:2009xs}, a threshold enhancement is observed
in the cross section around $W=3.9$ GeV, which provides an important
evidence for the role played by the $D\bar{D^*}+c.c.$ open
threshold. By fitting the data of Ref.~\cite{Abe:2006fj}, in
Ref.~\cite{Zhang:2009gy}, we show that the $\gamma^*D\bar{D^*}$ form
factor can be extracted based on the following two
parameterizations:

(i) Form factor one (FF-I) for the $\gamma^*D\bar{D^*}$ coupling:
\begin{equation}
g_{\gamma^* D \bar D^* }(s) = g_1 \exp{[-(s-(m_D+m_{D^*})^2)/t_1]} +
g_0 \ ,
\end{equation}
where $x=0$ corresponds to the $D\bar{D^*}+c.c.$ threshold, and
$g_1$, $t_1$, and $g_0$ are fitting parameters.

(ii) Form factor two (FF-II) for the $\gamma^*D\bar{D^*}$ coupling:
\begin{eqnarray}
g_{\gamma^* D \bar D^*}(s) = \left|\frac{b_0}{s - m_X ^2 + i m_X
\Gamma _X} + b_1\right| \ ,
\end{eqnarray}
with a background term $b_1$. The parameter $b_0$ can be regarded as
the product of the $\gamma^* X$ coupling and the $XD\bar{D^*}$
coupling.

\begin{figure}[h]
\begin{minipage}{14pc}
\includegraphics[width=14pc]{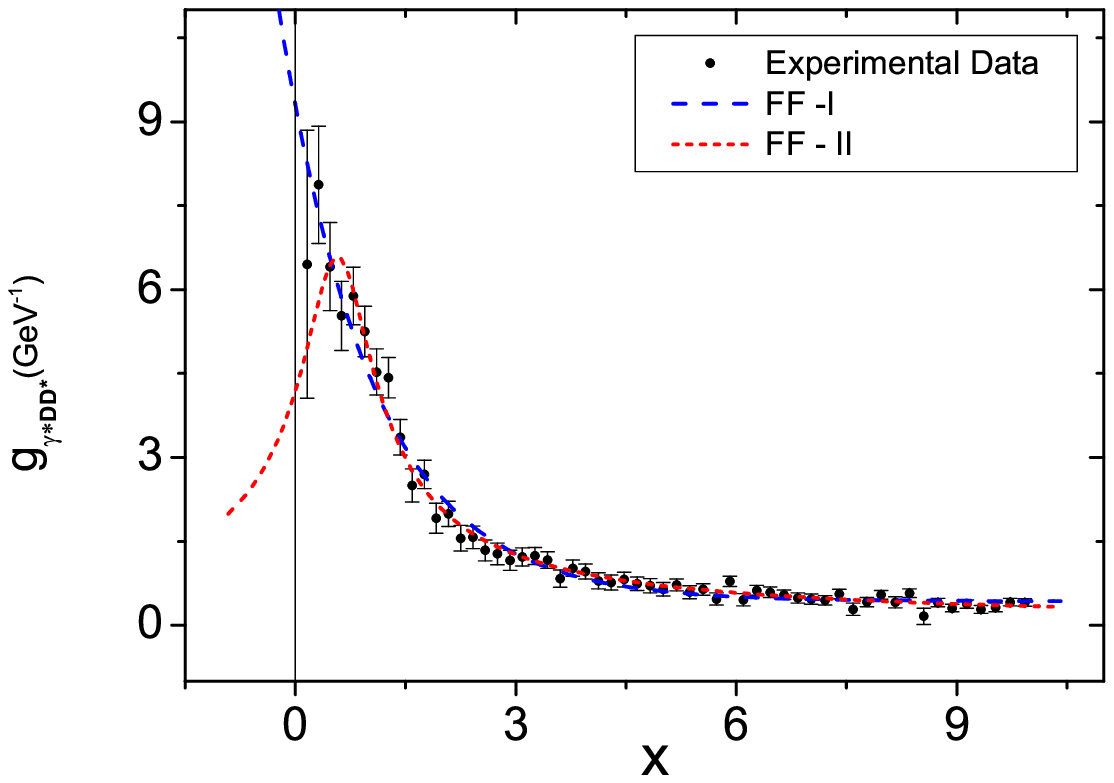}
\caption{\label{fig:1}The $x$-dependence of $g_{\gamma^* D\bar D^*}$
extracted from the cross sections for $e^+ e^- \to D^+ D^{*-} +
c.c.$ ~\cite{Abe:2006fj} with $x \equiv  s - (m_D + m_{D^*})^2$. The
dashed and dotted line are given by the FF-I and FF-II schemes,
respectively. The vertical line at $x=0$ labels the $D\bar{D^*}$
threshold.}
\end{minipage}\hspace{2pc}%
\begin{minipage}{14pc}
\includegraphics[width=14pc]{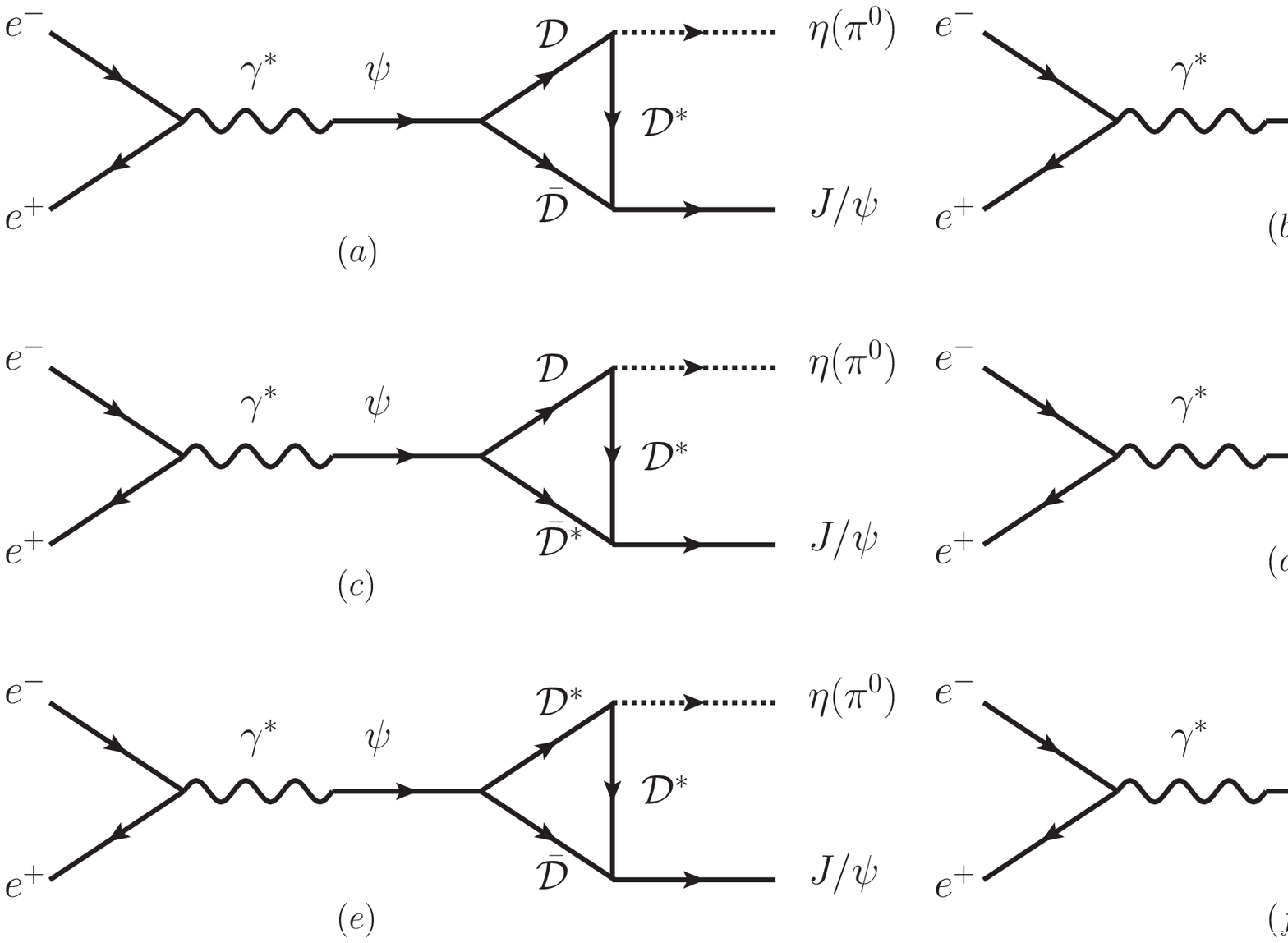}
\caption{\label{fig:2}Schematic diagrams for $e^+e^-\to
J/\psi\eta(\pi^0)$ via charmed $D$ ($D^*$) meson loops. The diagrams
for the $\phi\eta_c$ mode are similar.}
\end{minipage}
\end{figure}

The fitted parameters in these two parametrization schemes are
listed in Table.~\ref{ff-para}. As shown in Fig.~\ref{fig:1} by the
dashed line, FF-I can perfectly describe the energy dependence of
the effective coupling, and be extrapolated to the $D\bar{D}$
threshold. The dotted line indicates that the FF-II agrees with the
data at higher energies, but drops at the threshold region. In
Ref.~\cite{Zhang:2009gy}, it was shown that the structure of
$X(3900)$ in $e^+e^-\to D\bar{D}$ can be well described after taking
into account the $D\bar{D^*}+c.c.$ open threshold
effects~\cite{Zhang:2010zv}. However, one should realize that the
open threshold effects are not the unique explanation for the
structure of $X(3900)$. It can also be explained by a Breit-Wigner
distribution as shown by the parametrization of FF-II. So the key
question is, How would one be able to elucidate that the $X(3900)$
is caused by the $D\bar{D^*}+c.c.$ open threshold? The answer to
this question may possibly come from reaction channel $e^+e^-\to
J/\psi\pi^0$ as to be discussed later. Nevertheless, correlated
processes including $e^+e^-\to J/\psi\eta$ and $\phi\eta_c$ would
also bring additional information which can be examined in
experiment.

\begin{table}
\caption{ Parameters fitted with FF-I and FF-II for the $\gamma^*
D\bar{D^*}$ effective coupling form factor. The data are from
Belle~\cite{Abe:2006fj}.} \label{ff-para}
\begin{tabular}{|c|c|c|c|}\hline
\multicolumn{2}{|c|}{FF-I}
          & \multicolumn{2}{c|}{FF-II}\\\hline
\multicolumn{2}{|c|}{$~~~~g_{\gamma^* D \bar D^*}=g_1
\mbox{exp}[-(s-(m_D + m_{D^*})^2)/t_1] + g_0~~~~$} &
\multicolumn{2}{c|}{ $~~~~g_{\gamma^* D \bar D^*} = |\frac{b_0}{s -
m_X ^2 + i m_X \Gamma _X} + b_1|~~~~$}\\\hline $~~~~g_1~~~~$ & $8.86
\pm 0.59$ & $b_0$ & $3.08 \pm 0.31$
\\\hline
$t_1$ & $1.28 \pm 0.06$ & $m_X ~(\mbox{GeV})$ & $3.943 \pm 0.014$
\\\hline
$g_0$ & $0.43 \pm 0.02$ & $\Gamma_X ~(\mbox{MeV})$ & $119.0 \pm
10.0$
\\\hline
\multicolumn{2}{|c|}{} & $b_1$ & $0.016 \pm 0.045$
\\\hline
$~~~~\chi^2$/d.o.f ~~~~& 68.86/53  & $~~~~\chi^2$/d.o.f ~~~~  &
47.67/52
\\\hline
\end{tabular}
\end{table}

\subsection{Direct evidence for the $D\bar{D^*}+c.c.$ open threshold
effects in $e^+e^-\to J/\psi\pi^0$}

Recall that the position of the $X(3900)$ is located between the
known $\psi(3770)$ and $\psi(4040)$, its coupling to $J/\psi\eta$
and isospin-violating $J/\psi\pi^0$ would receive relatively small
interferences from the nearby resonances. Therefore, peculiar
threshold effects due to the open $D\bar{D^*}+c.c.$ might be
detectable in $e^+ e^-\to J/\psi\eta$ and $J/\psi\pi^0$ and
distinguish this structure from a resonance nature.  Apart from this
anticipation, we also consider the $\phi\eta_c$ channel, of which
the threshold is very close to the $D\bar{D^*}+c.c.$

As we know from the vector meson dominance (VMD) model, light vector
meson contributions to the cross sections are negligible in the
charmonium energy region. The main resonance contributions are from
vector charmonium excitations. Note that the final states $VP$
consist of a charmonium plus a light meson. Therefore, the
transitions are Okubo-Zweig-Iizuka (OZI) rule violating processes,
which should be dominated by soft mechanisms near threshold. Since
the pure electromagnetic (EM) transitions are negligibly small, a
natural way to recognize the soft mechanisms for $e^+e^-\to
J/\psi\eta$, $J/\psi\pi^0$ and $\phi\eta_c$ is via the open charm
transitions which are illustrated by Fig. \ref{fig:2}. Similar
approach has been applied to the study of the cross section
lineshape of $e^+e^-\to \omega\pi^0$ in the vicinity of the $\phi$
meson mass region~\cite{Li:2008xm}.

The effective Lagrangians for the coupling vertices involving
charmonia and charmed mesons are extracted from heavy quark
effective theory and chiral symmetry as applied in
Ref.~\cite{Liu:2009vv,Liu:2010um,Wang:2010iq}. They are written as
follows:
\begin{equation}
\mathcal{L}_2=i g_2 Tr[R_{c\bar{c}} \bar{H}_{2i}\gamma^\mu
{\stackrel{\leftrightarrow}{\partial}}_\mu \bar{H}_{1i}] + H.c.,
\end{equation}
where $R_{c\bar{c}}$ denotes the $S$-wave charmonium states, and
$H_{1i}$ and $H_{2i}$ are the  charmed and anti-charmed meson
triplet, respectively. The Lagrangian describing the interactions
between light meson and charmed mesons reads
\begin{eqnarray}
\mathcal{L}&=&iTr[H_iv^\mu \mathbf{D}_{\mu
ij}\bar{H}_j]+igTr[H_i\gamma_\mu\gamma_5A^\mu_{ij}\bar{H}_j]+i\beta
Tr[H_iv^\mu(V_\mu-\rho_\mu)_{ij}\bar{H}_j] \nonumber\\
&+&i\lambda Tr[H_i\sigma^{\mu\nu}F_{\mu\nu}(\rho)_{ij}\bar{H}_j],
\label{vph}
\end{eqnarray}
where the operator
$A_\mu=\frac{1}{2}(\xi^\dag\partial_\mu\xi-\xi\partial_\mu\xi^\dag)$
with $\xi=\sqrt{\Sigma}=e^{{iM}/{f_\pi}}$, and
$F_{\mu\nu}(\rho)\equiv
\partial_\mu\rho_\nu-\partial_\nu\rho_\mu+[\rho_\mu,\rho_\nu]$.
$M$ and  $\rho$ denote the light pseudoscalar octet and vector
nonet, respectively~\cite{Cheng:2004ru,Casalbuoni:1996pg}.

With the above Lagrangians, the leading order couplings are given as
follows~\cite{Cheng:2004ru}:
\begin{eqnarray}
\nonumber
 {\cal L} &=&-g_{\mathcal{D}^*\mathcal{D}\mathcal{P}}(\mathcal{D}^i\partial^\mu
\mathcal{P}_{ij}
 \mathcal{D}_\mu^{*j\dagger}+\mathcal{D}_\mu^{*i}\partial^\mu \mathcal{P}_{ij}\mathcal{D}^{j\dagger})
 +{1\over 2}g_{\mathcal{D}^*\mathcal{D}^*\mathcal{P}}
 \epsilon_{\mu\nu\alpha\beta}\,\mathcal{D}_i^{*\mu}\partial^\nu \mathcal{P}^{ij}
 {\stackrel{\leftrightarrow}{\partial}}{\!^\alpha} \mathcal{D}^{*\beta\dagger}_j
 \\\nonumber
 &-& ig_{\mathcal{D}\mathcal{D}\mathcal{V}} \mathcal{D}_i^\dagger {\stackrel{\leftrightarrow}{\partial}}{\!_\mu} \mathcal{D}^j(V^\mu)^i_j
 -2if_{\mathcal{D}^*\mathcal{D}\mathcal{V}} \epsilon_{\mu\nu\alpha\beta}
 (\partial^\mu \mathcal{V}^\nu)^i_j
 (\mathcal{D}_i^\dagger{\stackrel{\leftrightarrow}{\partial}}{\!^\alpha} \mathcal{D}^{*\beta j}+\mathcal{D}_i^{*\beta\dagger}{\stackrel{\leftrightarrow}{\partial}}{\!^\alpha} D^j)
 \\
 &+& ig_{\mathcal{D}^*\mathcal{D}^*\mathcal{V}} \mathcal{D}^{*\nu\dagger}_i {\stackrel{\leftrightarrow}{\partial}}{\!_\mu} \mathcal{D}^{*j}_\nu(\mathcal{V}^\mu)^i_j
 +4if_{\mathcal{D}^*\mathcal{D}^*\mathcal{V}} \mathcal{D}^{*\dagger}_{i\mu}(\partial^\mu \mathcal{V}^\nu-\partial^\nu
 \mathcal{V}^\mu)^i_j \mathcal{D}^{*j}_\nu,
 \label{eq:LDDV}
 \end{eqnarray}
where $\epsilon_{\alpha\beta\mu\nu}$ is the Levi-Civita tensor and
$D$ meson field destroys a $D$ meson. All the coupling values have
been given explicitly in Ref.~\cite{Wang:2011yh}. In this study, we
include five resonances, i.e. $J/\psi$, $\psi(3686)$, $\psi(3770)$,
$\psi(4040)$, and $\psi(4160)$ which are the dominant ones near the
$D\bar{D^*}+c.c.$ threshold.

As mentioned earlier, the reactions $e^+e^-\to J/\psi\eta$,
$J/\psi\pi^0$, and $\phi\eta_c$ via vector charmonium states
(denoted as $\Psi$) involve light hadron productions via soft gluon
exchanges. Therefore, they are generally OZI-rule suppressed. In
Refs.~\cite{Guo:2009wr,Guo:2010ak} it is shown that the intermediate
meson loops (IML) are relative enhanced in comparison with the soft
gluon exchanges for $\psi^\prime\to J/\psi\eta$ and $J/\psi\pi^0$.
As a consequence, the IML provide an mechanism to evade the OZI
rule. Nevertheless, it is also an important mechanism for causing
strong isospin violations in e.g. $\Psi\to J/\psi\pi^0$.

The reactions $e^+e^-\to J/\psi\eta$ and $J/\psi\pi^0$ are
correlated with each other as illustrated in Fig.~\ref{fig:2}.
Firstly, the meson loop amplitudes can be separated into two
classes, i.e. charged intermediate meson loops (CIML) and neutral
intermediate meson loops (NIML). In the isospin symmetry limit, i.e.
$m_u=m_d$, the absolute values of these two amplitudes are always
equal to each other. In the isospin conserved channels, such as
$\Psi\to J/\psi\eta$, these two amplitudes would add to each other,
while in the isospin-violating channels, such as $\Psi\to
J/\psi\pi^0$, these two amplitudes would have different signs and
thus cancel each other. In another word, in the isospin symmetry
limit, the cross section via the intermediate meson loops would
vanish in $e^+e^-\to \Psi\to J/\psi\pi^0$.

In reality, the isospin symmetry is broken, i.e. $m_u\neq m_d$.
Consequently, the masses of the charged and neutral $D \ (D^*)$
mesons are different. This will lead to nonvanishing cross sections
for $e^+e^-\to \Psi\to J/\psi\pi^0$ due to the incomplete
cancelations between the CIML and NIML amplitudes. As a prediction
for such a scenario, the residual cross sections will appear in
between the charged and neutral $D\bar{D^*}$ thresholds and produce
two cusps at the thresholds of $W=m_{D^0}+m_{\bar{D^{*0}}}=3871.79$
MeV and $W=m_{D^\pm}+m_{D^{*\mp}}=3879.85$ MeV. Namely, the
mechanism that causes the cross section enhancement between those
two thresholds is also responsible for the $X(3900)$ observed in
$e^+e^-\to D\bar{D}$~\cite{Pakhlova:2008zza}.

\begin{figure}[tb]
\begin{center}
\includegraphics[scale=0.45]{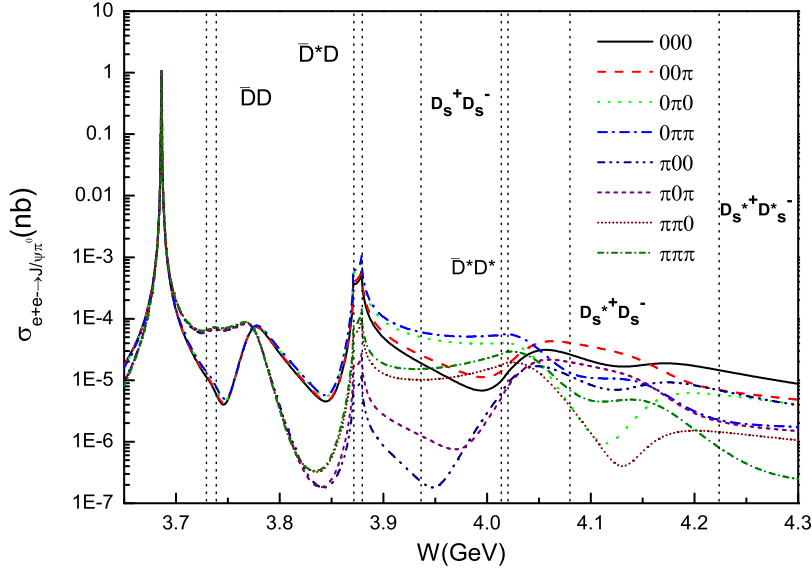}
\caption{The predicted cross section for $e^+e^-\to J/\psi\pi^0$ in
terms of the c.m. energy $W$ with different phase
angles~\protect\cite{Wang:2011yh}. The vertical lines labels the
open charm thresholds. }\label{fig:3}
\end{center}
\end{figure}

In Fig.~\ref{fig:3},  the calculated cross section  for $e^+ e^-\to
J/\psi\pi^0$  in terms of the c.m. energy $W$ is presented. Although
the cross section is rather sensitive to the relative phases
introduced among the transition amplitudes, the peak structure
$X(3900)$ has a model-independent feature and can be searched for in
experiment. More importantly, since the thresholds of
$D^0\bar{D}^{*0}+c.c.$ and $D^+D^{*-}+c.c.$ are isolated from the
known $\psi(3770)$ and $\psi(4040)$, the enhancement here would be a
clear evidence for non-resonant peaks in $e^+e^-$ annihilations. In
contrast, as shown in Ref.~\cite{Wang:2011yh}, although the
$D\bar{D}^*+c.c.$ loops have relatively large contributions to the
cross sections in $e^+e^-\to J/\psi\eta$, their contributions are
submerged by other amplitudes and cannot be indisputably identified
in the cross section lineshape.

We also note that  the $\psi(3686)$ has a predominant contribution
to $e^+e^-\to\Psi\to J/\psi\pi^0$  due to its strong isospin
violation couplings via the $D$ meson
loops~\cite{Guo:2009wr,Guo:2010ak}. Such a resonant enhancement
should be detectable, and the cross section measurement will provide
a calibration for the $X(3900)$ structure.

It should be pointed out that the $X(3900)$ as the open charm effect
is a collective one from the $D\bar{D}^*+c.c.$ loops to which all
the vector charmonia have contributions. That is why such a $P$ wave
configuration between $D\bar{D}^*+c.c.$ can produce the significant
enhancement in  $e^+e^-\to J/\psi\pi^0$. This mechanism is much
likely to be different from the $X(3872)$, which has been broadly
investigated in the literature as a dynamically generated
$D\bar{D}^*+c.c.$ bound state in the relative $S$ wave. Detailed
results for $e^+e^-\to J/\psi\eta$ and $\phi\eta_c$ can be found in
Ref.~\cite{Wang:2011yh}.

\section{Summary}

In summary, we have proposed to study the coupled channel effects in
$e^+e^-$ annihilating to $J/\psi\eta$, $J/\psi\pi^0$ and
$\phi\eta_c$.  In particular, we show that the reaction $e^+e^-\to
J/\psi\pi^0$ will be extremely interesting for disentangling the
resonance contributions and open charm effects taking the advantage
that the open $D\bar{D^*}$ threshold is relatively isolated from the
nearby known charmonia $\psi(3770)$ and $\psi(4040)$. Although we
also find that the predicted cross sections are rather sensitive to
the model parameters adopted, we clarify that the open charm effects
from the $D\bar{D^*}+c.c.$ channel are rather model-independent.
Therefore, it is extremely interesting to search for the predicted
enhancement around 3.876 GeV (i.e. $X(3900)$) in experiment.
Confirmation of this prediction would allow us to learn a lot about
the nature of non-pQCD in the charmonium energy region, and provide
insights into some of those long-standing puzzles in charmonium
decays, such as the ``$\rho$-$\pi$ puzzle" and $\psi(3770)$
non-$D\bar{D}$ decays etc~\cite{zhao-npb}.

\section*{Acknowledgments} Author wishes to acknowledge
collaborations with Q. Wang, X.-H. Liu, G. Li and Y.-J. Zhang on the
relevant research topics reported in the proceeding. This work is
supported, in part, by the National Natural Science Foundation of
China (Grant No. 11035006), the Chinese Academy of Sciences
(KJCX2-EW-N01), the Ministry of Science and Technology of China
(2009CB825200).

\end{document}